\documentclass{appolb}
\usepackage{graphicx}
\usepackage{enumerate}

\begin{document}
\title{Radiative origin of neutrino masses%
  \thanks{Presented at the XXXIX International Conference of
    Theoretical Physics ``Matter to the Deepest'', Ustro\'n, Poland,
    September 13-18, 2015.}%
}
\author{D. Aristizabal Sierra
\address{IFPA, Dep. AGO, Universit\'e de Li\`ege, Bat B5, Sart
  Tilman B-4000 Li\`ege 1, Belgium}
}
\maketitle
\begin{abstract}
  Mechanisms for Majorana neutrino mass generation can be classified
  according to the level at which the Weinberg operator is generated.
  The different possibilities can be sorted in ``canonical'' tree
  level and loop-induced realizations, the latter being motivated by
  their potential experimental testability.  Here we discuss the one-
  and two-loop cases, paying special attention to systematic
  classification schemes whose aim is that of constructing a full
  picture for neutrino mass generation.
\end{abstract}
  
\section{Introduction}
Neutrino oscillation experimental data has provided unquestionably
evidence for beyond-the-Standard-Model physics. Non-vanishing neutrino
mixing angles and neutrino masses require new degrees of freedom (dof)
whose scale is to a large extent a free parameter. An effective
description of the neutrino oscillation phenomena can be well
accounted for through the dimension-five effective operator
\begin{equation}
  \label{eq:weinberg-op}
  \mathcal{O}_5\sim \frac{C_{ij}}{\Lambda}\ell_i\,\ell_j\,H\,H\, ,
\end{equation}
the so-called Weinberg operator \cite{Weinberg:1980bf}. Pinning down
the origin of neutrino masses and mixings (and of CP violation),
however, requires unraveling the nature of the UV completion
responsible for this operator. The presence of new dof enable writing
the operator in a particular way, so different UV completions lead to
different realizations of $\mathcal{O}_5$. Though it might be as well
that the UV completion does not allow for $\mathcal{O}_5$, case in
which the resulting neutrino mass matrix will be determined by a
higher-order lepton-number-violating operator, see
e.g. \cite{Bonnet:2009ej}.

\section{Different forms of the Weinberg operator}
\label{sec:weinberg-op}
The conventional wisdom is that the new dof have masses at about the
GUT scale. The motivation for such ``belief'' resides on two
observations. First of all, order one couplings in
(\ref{eq:weinberg-op}), combined with an order 0.1 eV neutrino mass
\cite{Forero:2014bxa} fixes the lepton-number-violating scale at
$\Lambda\sim 10^{14}\,$~GeV. Secondly, such scale is very suggestive
of a new fundamental scale, which can be associated with a
GUT. Type-I, II and III seesaws fit perfectly within this paradigm,
and so can be regarded as ``orthodox'' approaches.

It can be as well that rather than being $\mathcal{O}\sim 1$, the
coupling in (\ref{eq:weinberg-op}) is smaller, thus implying smaller
$\Lambda$. Sticking to $\mathcal{O}_5$, two generic mechanisms can be
envisaged:
\begin{itemize}
\item The operator is generated radiatively. The suppression of the
  loop factors and extra couplings account for the smallness of $C$,
  thus assuring a smaller lepton-number-violating scale.
\item The operator involves small parameters whose values ``measure''
  the amount of lepton number breaking. These realizations, although
  involving the same UV completions that those of the ``orthodox''
  approaches, allow for smaller lepton-number-violating scales. 
\end{itemize}
In both categories the list of particular realizations is large, so
presenting here a complete listing is impossible. Focusing on the more
well-known cases, one can certainly argue that in the first category
at the one- and two-loop order the Zee and the Cheng-Li-Babu-Zee
models stand as the ``benchmark'' references
\cite{Zee:1980ai,Cheng:1980qt,Zee:1985id,Babu:1988ki} \footnote{Note
  that type-I seesaws where the neutrino mass matrix involves as well
  one-loop finite terms could be placed in this category, see
  \cite{Pilaftsis:1991ug,Grimus:1989pu,AristizabalSierra:2011mn}.}. This
is probably the reason why these cases have been the subject of
extensive phenomenological studies (see
e.g. \cite{AristizabalSierra:2006ri,AristizabalSierra:2006gb}).  Other
known examples involve colored scalars at the one- or two-loop level
\cite{AristizabalSierra:2007nf,Babu:2010vp} and radiative seesaw
realizations \cite{Ma:2006km}. The latter being as well subject to
throughout phenomenological analysis (see e.g. \cite{Sierra:2008wj}).
The inverse seesaw, on the other hand, is an example of a model where
the neutrino mass matrix involves extra suppression factors accounting
for slightly broken lepton number
\cite{Mohapatra:1986bd,Akhmedov:1995ip}.

Once a particular neutrino mass generating framework is fixed, the
origin of neutrino mixing can be addressed in several ways. The
``standard'' approach, however, relies on the idea that the UV
completion involves, in addition to the new dof, extra symmetries
which enforce the observed mixing pattern, see
e.g. \cite{King:2015aea,Morisi:2012fg}. Another approach, pointed out
in the context of the tribimaximal (TBM) pattern, consist of
``hybrid'' neutrino masses. Where the TBM structure is sourced by one
mechanism, while the experimentally required deviations by a another
one, see refs. \cite{Sierra:2013ypa} for more details.
\section{Sorting TeV-scale models systematically}
Although intrinsically useful, analyses based on particular models are
limited. Getting a more complete picture requires a systematical
treatment of different categories sharing common features. Starting
with the tree level \cite{Ma:1998dn}, systematic analyses of different
forms of the Weinberg operator (and higher lepton-number-breaking
operators too) have been pointed out. Two approaches have been
adopted: effective operator
\cite{Babu:2001ex,Choi:2002bb,deGouvea:2007xp,delAguila:2012nu,Angel:2012ug}
and diagrammatic classifications \cite{Bonnet:2012kz,Sierra:2014rxa},
with ``ingredients and recipes'' written up to the three-loop order
\cite{Farzan:2012ev}. In the diagrammatic case, classifications of
$\mathcal{O}_5$ are based on: $(a)$ identification of relevant
inequivalent renormalizable topologies (at a given order), $(b)$
systematic construction of relevant diagrams, $(c)$ quantum number
assignments, $(d)$ loop integrals calculation. Of fundamental
relevance in the overall classification, is the certainty that a
particle content of a given $n$-loop diagram does not generate a
leading $n-1$ (or below) contribution. This turns out to be
particularly relevant in the two-loop case, where such ``genuineness''
is assured by conditions placed over the possible particle content.
\begin{figure}
  \centering
  \includegraphics[scale=0.6]{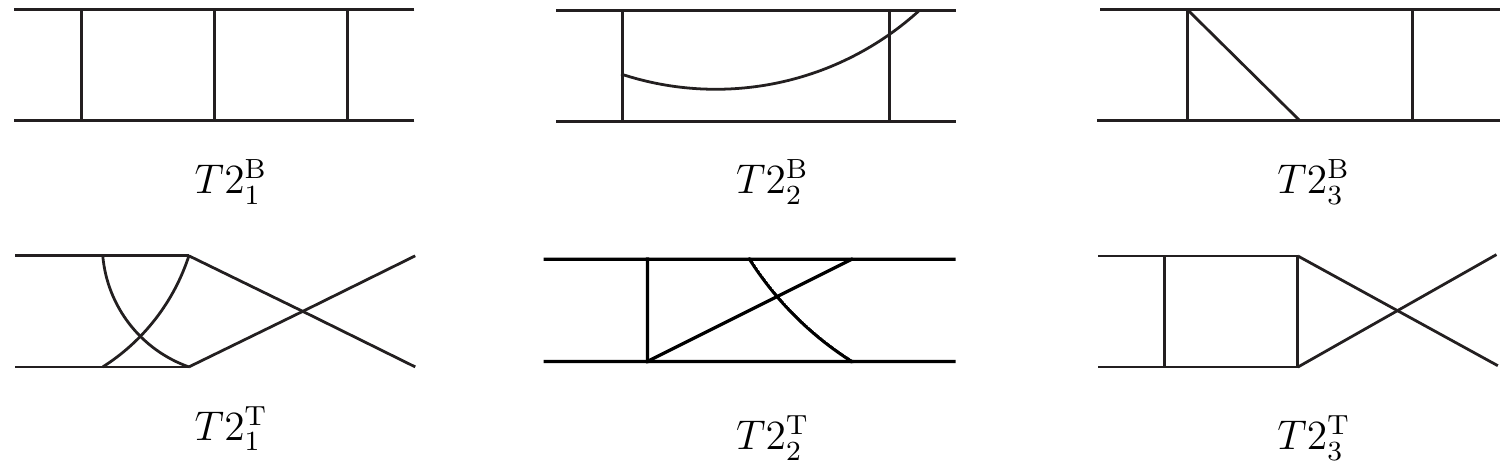}
  \caption{Relevant topologies for the two-loop Weinberg operator}
  \label{fig:relevant-topo}
\end{figure}
\subsection{Two-loop-induced neutrino mass models}
In the two-loop case one finds in total 29 topologies out of which
only six are relevant, as displayed in fig. \ref{fig:relevant-topo}
\cite{Sierra:2014rxa}. They are relevant in the following sense. The
different diagrams one can construct from the full set of topologies
can be sorted in three categories:
\begin{enumerate}[(I)]
\item \textit{Genuine diagrams}: Defined as diagrams for which the
  absence of one-loop and tree level realizations of $\mathcal{O}_5$
  is guaranteed. These diagrams are those that one can regard as
  leading to genuine two-loop models.
\item\textit{Non-genuine but finite diagrams}: These diagrams involve
  finite two-loop integrals. They define effective two-loop models:
  The neutrino mass matrix is one-loop-generated, but one of the
  ``inner'' couplings is generated radiatively at the one-loop order.
\item\textit{Non-genuine and divergent diagrams}: The two-loop
  integrals for these diagrams are divergent. Thus, all of them are
  ``just'' two-loop corrections to either tree or one-loop neutrino
  mass matrices. From that point of view, therefore, they are of no
  interest.
\end{enumerate}
Genuine diagrams arise from, and only from, the topologies shown in
fig.~\ref{fig:relevant-topo}, is in that sense that these topologies
are relevant. They however can as well generate non-genuine diagrams
falling into categories (II) and (III). Thus, they need to be endowed
with further rules (``genuineness rules'') that guarantee genuineness,
namely \cite{Sierra:2014rxa}: $(i)$ absence of hypercharge zero
fermion $SU(2)$ singlets and triplets or hypercharge two scalar
$SU(2)$ triplets; $(ii)$ absence of hypercharge zero scalar EW
singlets or triplets; $(iii)$ internal scalars should not have quantum
numbers matching those of the standard model Higgs ($H$); $(iv)$ for
quartic scalar couplings $H\,H\,S_1\,S_2$, the following choices:
$S_{1,2}=S_D$, $S_1=S_S$ and $S_2=S_T$, $S_1=S_T$ and $S_2=S_T$ (with
$S,D,T$ referring to singlet, doublet and triplet), require the
difference in hypercharge of these states to be different from $2Y_H$
($Y_H$ being the Higgs hypercharge).

\begin{figure}
  \centering
  \includegraphics[scale=0.6]{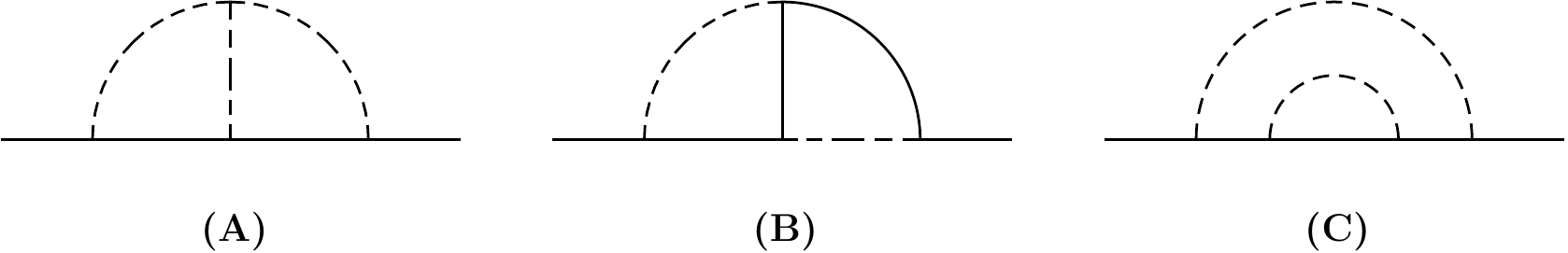}
  \caption{The three different classes of genuine two-loop
    diagrams. The different models arising from these diagrams depend
    on how the Higgs external legs are attached.}
  \label{fig:genuine-diagrams}
\end{figure}
\begin{figure}
  \centering
  \includegraphics[scale=0.65]{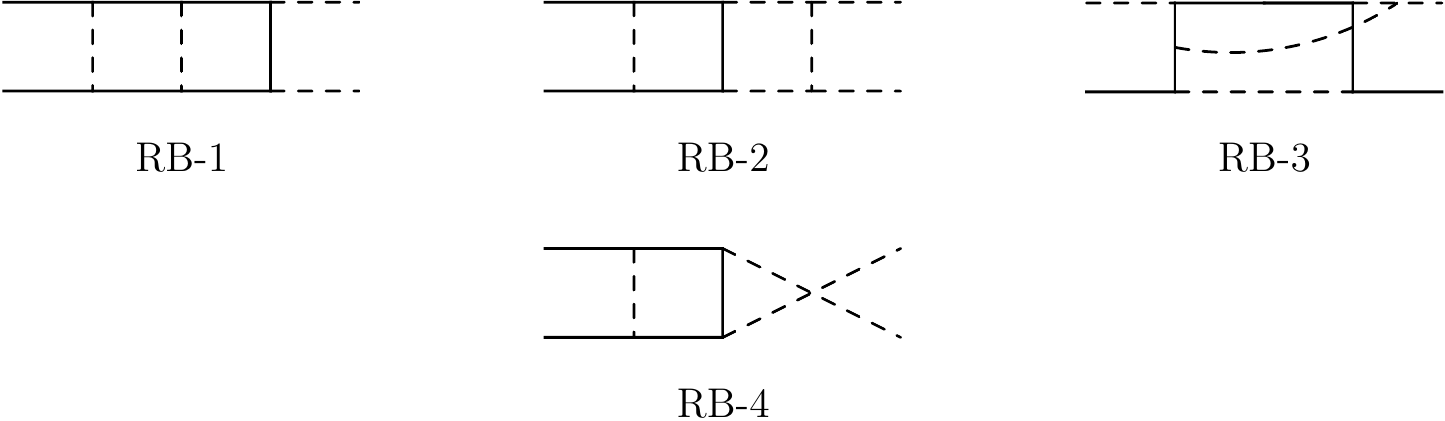}
  \caption{Class \textbf{(C)} possible diagrams.}
  \label{fig:RB-diag}
\end{figure}

The topologies in fig. \ref{fig:relevant-topo}, combined with the
above rules lead to a limited number of genuine diagrams falling in
three different non-overlapping classes, as shown in
fig. \ref{fig:genuine-diagrams}. The different genuine two-loop models
one can get are determined by the different ways in which the Higgs
external legs can be attached to a given diagram. According to the
different possibilities, ``genuineness rules'' allow for a quite
limited number of diagrams as follows: 10 diagrams for class
\textbf{(A)}, 6 diagrams for class \textbf{(B)} and 4 for class
\textbf{(C)}. Fig. \ref{fig:RB-diag} shows the four different diagrams
for class \textbf{(C)}.

Once the genuine diagrams are identified, $SU(2)$ quantum numbers of
the new fields are fixed by means of direct product decomposition. Due
to the two-loop character of the different diagrams, hypercharge is
determined up to two arbitrary constants. Sticking to lower EW
representations (singlets, doublets and triplets), all possible
$SU(2)\times U(1)_Y$ quantum numbers have been presented in
\cite{Sierra:2014rxa}. These results, along with tabulated two-loop
integrals presented as well in \cite{Sierra:2014rxa}, provide a
complete catalog for radiative neutrino masses at the two-loop
order.
\section{Conclusions}
\label{sec:concl}
Among the several mechanism one can envisage for neutrino mass
generation, radiatively-induced neutrino masses are a well-motivated
option. Certainly, their motivation resides on the fact that in
contrast to the ``standard'' tree level mechanism, loop-induced
neutrino masses are (in principle) testable. Here, after sketching the
different pathways that can be considered (beyond the tree level and
radiative cases), we have discussed the different possibilities for
the two-loop case. The results presented here, along with the results
from the one-loop systematic classification, complete the model
building picture for radiative neutrino masses up to the two-loop
order.
\section*{Acknowledgements}
I would like to thank Audrey Degee, Luis Dorame, Diego Restrepo and
Carlos Yaguna for the collaboration on some of the subjects discussed
here. I would like to specially thank Martin Hirsch for the
collaboration that led to some of the papers quoted here as well as
for the always useful conversations.

\end{document}